\documentclass[fleqn,usenatbib]{mnras}

\usepackage{newtxtext,newtxmath}

\usepackage[T1]{fontenc}

\DeclareRobustCommand{\VAN}[3]{#2}
\let\VANthebibliography\thebibliography
\def\thebibliography{\DeclareRobustCommand{\VAN}[3]{##3}\VANthebibliography}

\usepackage{graphicx}	
\usepackage{amsmath}	
\usepackage{float}
\usepackage{booktabs}

\title[HESS J1825$-$137 and its environment]
{Constraining leptonic and hadronic $\gamma$-ray emission 
from HESS~J1825$-$137 and its environment}

\author[Rubens Costa Jr. and R. C. Anjos]{
Rubens Costa Jr.,$^{1}$\thanks{E-mail: rubensp@utfpr.edu.br}
R. C. Anjos,$^{1,2,3,4,5}$\thanks{E-mail: ritacassia@ufpr.br}
\\
$^{1}$Programa de Pós-Graduação em Física e Astronomia, Universidade Tecnológica Federal do Paraná, Curitiba, PR 80230-901, Brazil\\
$^{2}$Departamento de Engenharias e Exatas, Universidade Federal do Paraná (UFPR), Pioneiro, 2153, 85950-000 Palotina, PR, Brazil\\
$^{3}$N\'ucleo de Astrof\'{\i}sica e Cosmologia (Cosmo-Ufes) \& Departamento de F\'isica, Universidade Federal do Esp\'irito Santo, 29075--910, Vit\'oria, ES, Brazil\\
$^{4}$Programa de pós-graduação em Física, Universidade Estadual de Londrina (UEL), Rodovia Celso Garcia Cid Km 380, 86057-970 Londrina, PR, Brazil\\
$^{5}$Programa de Pós-Graduação em Física Aplicada, Universidade Federal da Integração Latino-Americana, 85867-670, Foz do Igua\c{c}u, PR, Brazil\\
}

\date{Accepted 2026 June 11. Received 2026 March 26; in original form 2026 March 26}
\pubyear{\the\year{}}

\begin{document}
\label{firstpage}
\pagerange{\pageref{firstpage}--\pageref{lastpage}}
\maketitle

\begin{abstract}
We present a broadband spectral analysis of the $\gamma$-ray 
emission from the pulsar wind nebula HESS~J1825$-$137, 
combining observations from Fermi Large Area Telescope (\textit{Fermi}-LAT), High Energy Stereoscopic System (H.E.S.S.), High-Altitude Water Cherenkov Observatory (HAWC), and Very Energetic Radiation Imaging Telescope Array System (VERITAS) across the $\sim 0.1$~GeV--$160$~TeV 
energy range. The spectral energy distribution is modelled 
under purely leptonic, purely hadronic, and lepto-hadronic scenarios using the \textsc{Naima} 
radiative modeling framework with Markov Chain Monte Carlo parameter 
estimation. Model comparison via the Bayesian Information Criterion 
reveals that the baseline GeV--TeV data favour a purely 
leptonic interpretation, while the inclusion of simulated 
Cherenkov Telescope Array Observatory (CTAO) observations or Large High Altitude Air Shower Observatory (LHAASO) ultra-high-energy (UHE; $E_{\gamma} \ge 100\,\mathrm{TeV}$) measurements 
shifts the preference toward models incorporating a hadronic 
component ($\Delta\mathrm{BIC} = -28.87$ and $-7.89$, 
respectively). The inferred electron energy budget for the baseline GeV--TeV dataset,
$W_e = 4.25 \times 10^{48}$~erg, is consistent with previous estimates
reported in the literature. The 
proton energy budget, 
$W_p \approx 2.5 \times 10^{48}$~erg, is 
energetically compatible with $pp$ interactions in the dense 
molecular environment adjacent to the nebula. These results 
demonstrate that precise spectral measurements above 
$\sim 10$~TeV, where Klein--Nishina suppression of inverse 
Compton emission creates a window for hadronic processes, are 
essential to establish the dominant emission mechanism in this 
source.
\end{abstract}



\section{Introduction}
\label{sec:introduction}

Ultrarelativistic particles are crucial in various astrophysical systems, including pulsars, supernova remnants (SNRs), active Galactic nuclei, and galaxy clusters. In the interstellar medium of our Galaxy, cosmic rays (CRs) are nearly in pressure equilibrium with the turbulent gas motion and magnetic fields~\citep{aharonian2005, adriani2011, bykov2013, Spurio2018}. Despite their dynamical importance, the interplay between these components and their impact on star formation and Galactic evolution remain poorly understood. A key feature of the CR energy spectrum is the so-called ``knee'' at energies around 1~PeV\footnote{1~PeV = $10^{15}$~eV.}, which is generally interpreted as evidence that CRs are predominantly of Galactic origin up to this energy~\citep{ANTONI20051}. Identifying astrophysical accelerators capable of reaching PeV energies, commonly referred to as PeVatrons, is therefore essential for understanding the origin of Galactic CRs~\citep{Gabici_2007, Albert_2020, Cao2021b, 2023AppSc..13.6433C, Acero_2023, Costa2024, Auger_2024}.

In recent years, a growing number of Galactic ultra-high-energy (UHE; $E_{\gamma} \ge 100\,\mathrm{TeV}$) gamma-ray sources have been reported by ground-based facilities, including imaging atmospheric Cherenkov telescopes such as the High Energy Stereoscopic System (H.E.S.S.) and the Very Energetic Radiation Imaging Telescope Array System (VERITAS), as well as wide-field air-shower arrays including the High-Altitude Water Cherenkov Observatory (HAWC), the Tibet Air Shower Array (Tibet~AS$_{\gamma}$), and the Large High Altitude Air Shower Observatory (LHAASO)~\citep{2020PhRvL.124b1102A, ABEYSEKARA2020102403, 2018A&A...612A...2H, Cao2021b, 2021PhRvL.127c1102A}. Among the various classes of Galactic very-high-energy (VHE, $E_{\gamma}\ge0.1\,$TeV) emitters, pulsar wind nebulae (PWNe) have emerged as one of the most prominent populations~\citep{2018A&A...612A...2H, 2022A&A...660A...8B}. PWNe form when the relativistic particle wind driven by the rotational energy of a young pulsar is decelerated at a termination shock upon interaction with the surrounding medium, typically the slowly expanding ejecta of the progenitor supernova~\citep{GaenslerSlane2006}. Their $\gamma$-ray emission is governed by the properties of the central neutron star, the relativistic outflow powered by its rotational energy, and the surrounding environment~\citep{2022hxga.book...61M}. Evolved PWNe are particularly valuable for studying particle transport and radiative cooling, as their large spatial extent can provide spatially resolved spectra under favourable circumstances~\citep{GaenslerSlane2006, 2020A&A...640A..76P}. Observational studies indicate that the most luminous PWNe are typically associated with young and energetic pulsars with spin-down luminosities ($\dot{E}>10^{35}$~erg~s$^{-1}$)~\citep{2011ApJ...726...35A, 2024arXiv240802070X}.

HESS~J1825--137, discovered in 2005 by the H.E.S.S. Collaboration, is one of the most prominent PWNe in the Galaxy. The source is associated with the energetic pulsar PSR~J1826--1334~\citep{Manchester2005}, also known as PSR~B1823--13~\citep{Clifton1986, Yuan2010, Yao_2017}. With an angular extent of approximately $1^{\circ}$, HESS~J1825--137 ranks among the largest resolved VHE $\gamma$-ray sources~\citep{2005A&A...442L..25A}, and its luminosity places it among the brightest Galactic emitters in this energy range~\citep{2019A&A...621A.116H, 2023AppSc..13.6433C}. A defining observational characteristic of the nebula is its strongly energy-dependent morphology, with the $\gamma$-ray spectrum softening progressively with increasing distance from the pulsar---a signature attributed to the radiative cooling of electrons as they are transported away from the acceleration site~\citep{2006A&A...460..365A, 2019A&A...621A.116H}. Owing to its exceptional size and brightness, the nebula has become a benchmark object for studies of particle transport, radiative cooling, and energy-dependent morphology in evolved PWNe.

At the highest energies, HESS~J1825--137 lies in a complex region of the Galactic plane populated by multiple extended $\gamma$-ray sources detected by wide-field instruments. The nebula is spatially coincident with the HAWC source 2HWC~J1825--138 and lies within $\sim1^{\circ}$ of the LHAASO source J1825--1326. Owing to the limited angular resolution of air-shower arrays, the origin of the observed PeV emission in this region remains uncertain, with nearby sources such as HESS~J1826--130 also proposed as potential contributors~\citep{Cao2021b, Costa2024}. In its first source catalogue, LHAASO further reported HESS~J1825--137 as the closest known counterpart to 1LHAASO~J1825--1418 and 1LHAASO~J1825--1337u~\citep{2024ApJS..271...25C}. These spatial overlaps underscore the need for broadband spectral modelling and careful source association when interpreting the GeV--to--PeV $\gamma$-ray emission from this region. A preliminary multi-instrument analysis of the 
LHAASO~J1825$-$1326 region was presented 
by~\citet{Costa2024}, who modelled the sub-regions 
associated with PSR~J1826$-$1256 and PSR~J1826$-$1334 using 
\textsc{GALPROP}~\citep{2022ApJS..262...30P}, finding a predominantly leptonic high-energy 
emission with hadronic contributions below $\sim\!1\%$. 
However, that study did not employ a radiative framework to 
self-consistently model the relevant energy-loss processes, 
nor a formal statistical criterion for model comparison.

In this work, we perform a comprehensive spectral analysis of HESS~J1825--137, modelling its $\gamma$-ray emission under purely leptonic, purely hadronic, and lepto-hadronic scenarios. We use multi-instrument data from  Large Area Telescope (\textit{Fermi}--LAT)~\citep{Abdollahi2022,ballet2024}, H.E.S.S.~\citep{2019A&A...621A.116H}, HAWC~\citep{Albert2021}, VERITAS~\citep{ABEYSEKARA2020102403}, and LHAASO~\citep{2024ApJS..271...25C}, adopting a Bayesian spectral modelling approach based on \textsc{Naima}~\citep{naima} to constrain the underlying particle populations. To quantitatively compare competing emission scenarios, we further apply model selection criteria based on the Bayesian Information Criterion (BIC; \citealt{Schwarz1978}). In addition, the expected response of the Cherenkov Telescope Array Observatory (CTAO)~\cite{2019scta.book.....C} is evaluated using simulated instrument response functions (IRFs) in order to assess the diagnostic power of next-generation observations~\citep{Costa2024, 2025ApJ...979...23S}.

Within this framework, we explicitly consider two complementary modelling configurations: a \emph{Baseline} model excluding LHAASO data, which provides an updated and self-consistent extension of the H.E.S.S. Collaboration results based on current GeV--TeV observations, and a \emph{UHE-extended} model including LHAASO measurements, which enables us to explore for the first time the impact of UHE constraints on the inferred particle distributions and emission mechanisms in HESS~J1825--137. The main goals of this study are to identify the dominant emission mechanisms across the GeV--TeV energy range, to quantify the relative contributions of leptonic, hadronic, and lepto-hadronic processes, and to explore the extent to which future CTAO observations can resolve current ambiguities in this complex region.

The paper is structured as follows. Section~\ref{sec:2} presents the observational datasets and environmental characteristics of the region. Section~\ref{sec:3} describes the spectral models and statistical framework. Section~\ref{sec:4} presents the results, while Section~\ref{sec:5} discusses their physical interpretation. Finally, Section~\ref{sec:6} summarises the main conclusions and outlines prospects for future CTAO observations.

\section{Multi-GeV to Sub-PeV $\gamma$-ray Observations}
\label{sec:2}

HESS~J1825$-$137 has been firmly detected across the 
GeV--TeV energy range by several $\gamma$-ray observatories, 
establishing it as a persistent and extended high-energy 
source associated with the PWN powered by 
PSR~J1826$-$1334. This section summarises the key properties 
of the pulsar and the multiwavelength characteristics of the 
nebula relevant for the spectral modelling, followed by a 
description of the $\gamma$-ray datasets used in the present 
analysis.

\subsection{The pulsar and its environment}
\label{sec:2.1}

PSR~J1826--1334 (PSR~B1823--13) was discovered in the Jodrell Bank 20~cm radio survey~\citep{Clifton1992}. It has a spin period of $P = 101.48$~ms, a characteristic age of $\tau_c = 21.4$~kyr, and a spin-down luminosity of $\dot{E}=2.8\times10^{36}$~erg~s$^{-1}$, placing it among the most energetic pulsars in the Australia Telescope National Facility (ATNF) Pulsar Catalogue and rendering it a typical Vela-like pulsar~\citep{Manchester2005}. The distance to the pulsar, $d=3.9\pm0.4$~kpc, is inferred from its dispersion measure ($\mathrm{DM}=231\,\mathrm{pc\,cm}^{-3}$) using the NE2001 Galactic electron-density model~\citep{CordesLazio2002}. Despite its young age, deep Karl G. Jansky Very Large Array (VLA) radio observations have not detected an associated SNR~\citep{Braun1989, Gaensler2000}.

The nebula was first detected in X-rays by \textit{ROSAT}, which revealed a compact structure of $\sim20''$ radius around the pulsar~\citep{1996ApJ...466..938F}. \textit{XMM-Newton} observations subsequently identified both a compact core ($\sim30''$) and an asymmetric diffuse component extending $\sim5'$ to the south, attributed to synchrotron emission from the PWN~\citep{2003ApJ...588..441G}. \textit{Chandra} resolved the innermost PWN structure, revealing a compact nebula of $\sim25'' \times 10''$ with a hard spectral index 
($\Gamma_X \approx 1.3$) steepening to $\Gamma_X \approx 1.9$ in the 
surrounding diffuse emission extending at least $2.4'$ to the 
south~\citep{Pavlov2008}. \textit{Suzaku} extended the detection to $15'$ ($\sim17$~pc) from the pulsar, with a photon index of $\Gamma_X \approx 2.0$ in the diffuse component and a magnetic field of $\sim7\,\mu$G inferred from the combined X-ray/$\gamma$-ray spectral energy distribution~\citep{Uchiyama2009}. In the radio band, VLA observations at 1.4~GHz detected a compact source centred on the pulsar position, but no extended emission comparable to the $\gamma$-ray extent was found~\citep{Duvidovich2019}.

The asymmetric morphology of the nebula, extending predominantly to the south of the pulsar, is attributed to the interaction of the expanding SNR with dense molecular material to the north, as revealed by CO and CS observations~\citep{2016MNRAS.458.2813V}. This dense medium is thought to have produced an asymmetric reverse shock that compressed the nebula preferentially towards the south~\citep{2003ApJ...588..441G, 2006A&A...460..365A}. The presence of molecular clouds with hydrogen densities of $\sim400$--$600\,\mathrm{cm}^{-3}$ in the vicinity also provides potential targets for hadronic $\gamma$-ray production~\citep{2016MNRAS.458.2813V}.

Deep H.E.S.S. observations, combining more than twelve years of H.E.S.S.~I data with H.E.S.S.~II observations, revealed that the nebula emission extends out to $1.5^{\circ}$ from the pulsar, yielding an intrinsic diameter of $\sim100$~pc and making it potentially the largest $\gamma$-ray PWN currently known~\citep{2019A&A...621A.116H}. The energy dependence of the nebula extent follows $R \propto E^{\alpha}$ with $\alpha = -0.29 \pm 0.04_\mathrm{stat} \pm 0.05_\mathrm{sys}$, which disfavours a purely diffusive transport scenario~\citep{2019A&A...621A.116H}. At GeV energies, the first detailed \textit{Fermi}-LAT morphological study by \cite{2020A&A...640A..76P}, using 11.6~years of data between 1~GeV and 1~TeV, showed that the emission region extends beyond $2^{\circ}$ ($\sim150$~pc), with the nebula radius continuing to increase toward lower energies.

Several particle transport models have been developed for this source. \cite{VanEtten2011} performed time-dependent multi-zone modelling, finding that a combination of advective transport with a radially decreasing velocity profile ($v(r) \propto r^{-0.5}$) and a true age of $40 \pm 9$~kyr reproduces the multiwavelength data. \cite{2024MNRAS.528.2749C} presented a 3D diffusive and advective model, confirming that an older age ($\sim40$~kyr) and an advective flow ($v = 0.002c$) are needed, and that a turbulent ISM with magnetic fields of $20$--$60\,\mu$G to the north is required to prevent $\gamma$-ray contamination towards HESS~J1826--130. More recently, \cite{2024A&A...690A.116M} developed a multi-zone framework accounting for particle escape from the PWN into the SNR and ISM, finding that the extended GeV--TeV emission can be largely attributed to escaped particles, while the more compact emission above $50$--$100$~TeV originates within the PWN itself.

\subsection{GeV--TeV observations}
\label{sec:2.2}

In the GeV energy range, the source has been consistently reported as an extended emitter in the \textit{Fermi}--LAT catalogues, appearing as 3FGL~J1824.5$-$1351e~\citep{Acero_2015}, 3FHL~J1824.5$-$1351e \citep{Ajello2017}, and most recently as 4FGL~J1824.4$-$1350e in the fourth \textit{Fermi}--LAT catalogue~\citep{Abdollahi2022, ballet2024}. These successive detections demonstrate the robustness of the GeV emission over more than a decade of observations.

At TeV energies, HESS~J1825--137 is among the brightest and most extended PWNe in the Galaxy. Extensive observations by the H.E.S.S. Collaboration have yielded a well-constrained spectral energy distribution from multi-GeV to multi-TeV energies based on more than twelve years of H.E.S.S. observations, complemented by \textit{Fermi}-LAT data~\citep{2019A&A...621A.116H}. The total $\gamma$-ray flux above 1~TeV is $(1.12 \pm 0.03_\mathrm{stat} \pm 0.25_\mathrm{sys}) \times 10^{-11}\,\mathrm{cm}^{-2}\,\mathrm{s}^{-1}$, corresponding to approximately 64\% of the Crab Nebula flux~\citep{2019A&A...621A.116H}. At higher energies, \cite{Albert2021} resolved the previously 
blended HAWC emission from this region into three components, 
identifying HAWC~J1825$-$138 as the counterpart of 
HESS~J1825$-$137. The spectrum of this component exhibits 
significant curvature incompatible with a simple power law, 
constraining the maximum energy of the electron population 
powering the PWN. The same analysis also revealed a new hard 
source, HAWC~J1825$-$134, whose emission extends beyond 
$200\,\mathrm{TeV}$ without a cutoff.

Independent observations by VERITAS further confirm the detection of the nebula at TeV energies, with a reported statistical significance of $6.7\sigma$ and a flux above 1~TeV of $(3.9 \pm 0.8)\times10^{-12}\,\mathrm{cm^{-2}\,s^{-1}}$~\citep{ABEYSEKARA2020102403}. Taken together, the GeV and TeV measurements provide a continuous view of the $\gamma$-ray emission from HESS~J1825$-$137 over more than four decades in energy, with VERITAS providing independent constraints on the TeV flux normalisation and spectral shape.

The combined GeV--TeV spectral energy distribution (SED), constructed from \textit{Fermi}--LAT~\citep{Abdollahi2022,ballet2024}, H.E.S.S.~ \citep{2019A&A...621A.116H}, HAWC~\citep{Albert2021}, and VERITAS~\citep{ABEYSEKARA2020102403} measurements, constitutes the \emph{Baseline} dataset used in the spectral modelling presented in this work. The GeV--TeV emission of the nebula has been consistently characterised across successive \textit{Fermi}--LAT catalogues~\citep{Acero_2015, Ajello2017}, while combined H.E.S.S.\ and \textit{Fermi}--LAT analyses 
have investigated particle transport within the 
nebula~\citep{2019A&A...621A.116H}.

\subsection{Ultra-high-energy constraints}
\label{sec:2.3}

At the highest energies, the region surrounding HESS~J1825--137 has been probed by LHAASO. The source lies within $1^{\circ}$ of the LHAASO detection J1825--1326~\citep{Cao2021b, Costa2024}, and is identified as the closest known counterpart to 1LHAASO~J1825--1418 (separation $0.56^{\circ}$) and 1LHAASO~J1825--1337u ($0.15^{\circ}$) in the first LHAASO source catalogue~\citep{2024ApJS..271...25C}. Due to the limited angular resolution of air-shower arrays, the association between the PeV emission and HESS~J1825--137 remains uncertain, with nearby sources such as HESS~J1826--130 also proposed as potential contributors~\citep{Cao2021b, Costa2024}. HESS~J1826--130 was originally considered an extension of HESS~J1825--137 until it was resolved as a separate hard-spectrum source in the H.E.S.S. Galactic Plane Survey~\citep{2018A&A...612A...2H}, and has since been associated with the Eel PWN (G18.5--0.4) and PSR~J1826--1256~\citep{Duvidovich2019, Anguner2017}.

\begin{figure*}
    \centering
    \includegraphics[width=0.7\textwidth]{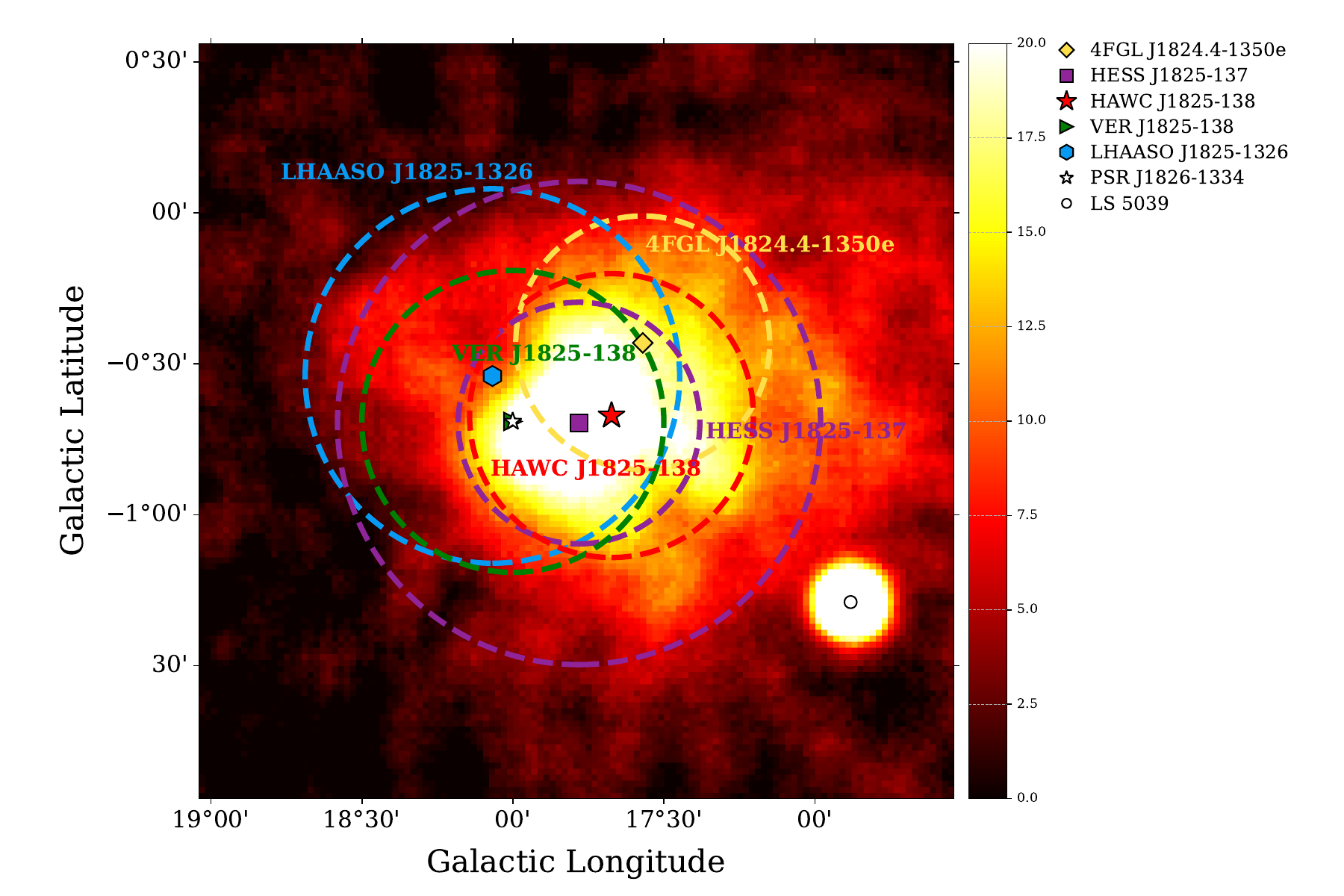}
    \caption{H.E.S.S.  significance skymap of the region surrounding HESS~J1825--137, 
    derived from the H.E.S.S.  Galactic Plane Survey (HGPS) \citep{Abdalla2018}, 
    shown in Galactic coordinates. 
    The background intensity scale indicates the detection significance in units of standard 
    deviations ($\sigma$).
    The positions of multi-wavelength counterparts and neighbouring sources are 
    overlaid: 
    HESS~J1825--137~\citep{2019A&A...621A.116H}, 
4FGL~J1824.4--1350e~\citep{Abdollahi2022,ballet2024}, 
HAWC~J1825--138~\citep{Albert2021}, 
VER~J1825--138~\citep{ABEYSEKARA2020102403}, 
and LHAASO~J1825--1326~\citep{2024ApJS..271...25C}. 
The associated pulsar PSR~J1826--1334 \citep{Manchester2005} is marked 
with a star at $(l,b) = (18.00^\circ, -0.69^\circ)$, 
and the gamma-ray binary LS~5039 \citep{Aharonian2006} 
with an open circle at $(l,b) = (16.88^\circ, -1.29^\circ)$.
Dashed circles indicate the reported spatial extensions of the different gamma-ray sources. For HESS~J1825--137, two characteristic regions are 
shown, corresponding to its core emission and its larger-scale nebular extent.
    } \label{fig:skymap}
\end{figure*}

Figure~\ref{fig:skymap} highlights the spatial distribution of the very-high-energy $\gamma$-ray emission in the vicinity of HESS~J1825$-$137, showing its extended morphology and the relative positions of nearby GeV--PeV sources detected by \textit{Fermi}-LAT, HAWC, VERITAS, and LHAASO, as well as the associated pulsar PSR~J1826$-$1334 and the $\gamma$-ray binary LS~5039.

\subsection{Datasets and analysis strategy}
\label{sec:2.4}

We analyse two datasets: a \emph{Baseline} dataset comprising 
\textit{Fermi}--LAT~\citep{Abdollahi2022,ballet2024}, 
H.E.S.S.\ \citep{2019A&A...621A.116H}, 
HAWC~\citep{Albert2021}, 
and VERITAS~\citep{ABEYSEKARA2020102403} measurements, 
and an \emph{UHE-extended} dataset that supplements these with 
LHAASO observations~\citep{2024ApJS..271...25C}. 
These datasets provide the foundation for the spectral modelling 
and model-comparison analyses presented below, and allow us to 
quantify the impact of sub-PeV to PeV measurements on the 
inferred emission scenarios. All models are fitted independently to each dataset to assess the effect of the LHAASO data on model preference and physical parameters. Section~\ref{sec:3} introduces the spectral modelling approach adopted in this work, encompassing inverse-Compton (IC), pion-decay (PD), and leptohadronic (LH) emission scenarios.

\section{Spectral modelling framework}
\label{sec:3}

This section describes the emission models and statistical methods 
used to interpret the broadband GeV--TeV $\gamma$-ray spectrum of 
HESS~J1825$-$137. All spectral modelling is performed with 
\textsc{Naima}~\citep{naima}, and parameter estimation 
is carried out within a Markov Chain Monte Carlo (MCMC) 
framework~\citep{MacKay2004, MacKay2013}. The analysis uses the 
combined SED constructed from \textit{Fermi}-LAT, H.E.S.S., HAWC, 
and VERITAS data described in Sect.~\ref{sec:2}. To assess the 
constraining power of next-generation facilities, simulated CTAO 
observations are also included, produced with the \textsc{Gammapy} 
analysis toolkit~\citep{Donath_2023} and CTAO instrument response 
functions (prod5 v0.1; \citealt{CTAIRFS2021}).

The leptonic $\gamma$-ray emission is modelled as IC scattering of relativistic electrons on ambient photon fields. 
We adopt the interstellar radiation field parameters used by 
\citet{2019A&A...621A.116H}, based on the Galactic model of 
\citet{Popescu2017}. This includes contributions from the cosmic 
microwave background (CMB; $T = 2.7$~K, 
$\omega = 0.25\,\mathrm{eV\,cm}^{-3}$), far-infrared (FIR; 
$T \sim 40$~K, $\omega \sim 1\,\mathrm{eV\,cm}^{-3}$), near-infrared 
(NIR; $T \sim 500$~K, 
$\omega \sim 0.4\,\mathrm{eV\,cm}^{-3}$), and optical/visible (VIS; 
$T \sim 3500$~K, 
$\omega \sim 1.9\,\mathrm{eV\,cm}^{-3}$) photon fields.

The hadronic $\gamma$-ray emission arises from PD following 
inelastic proton--proton ($pp$) collisions. As target material, we 
adopt the hydrogen density inferred from CO(1--0) observations by 
\citet{2016MNRAS.458.2813V}, who identify molecular material 
spatially adjacent to HESS~J1825$-$137 at a distance consistent 
with that of PSR~J1826$-$1334. Assuming a spherical molecular cloud 
of radius $R_{\mathrm{MC}} \sim 18$~pc, the average hydrogen density 
is $n_{\mathrm{H}} \simeq 6 \times 10^{2}\,\mathrm{cm}^{-3}$. This 
value represents an effective local density associated with dense 
molecular clumps, as traced by CO(1--0) and CS(1--0) emission, 
rather than a volume-averaged density over the full extent of the 
nebula. Such dense substructures provide suitable targets for 
relativistic protons diffusing out of the PWN, enabling efficient 
PD $\gamma$-ray production in a spatially localised region.

The energy distributions of both electron and proton populations 
are described by a broken power law (BPL),
\begin{equation}
N(E) =
\begin{cases}
N_{0} \left(E/E_{0}\right)^{-\alpha_{1}}, & E < E_{\mathrm{break}}, \\
N_{0} \left(E_{\mathrm{break}}/E_{0}\right)^{\alpha_{2}-\alpha_{1}}
\left(E/E_{0}\right)^{-\alpha_{2}}, & E \ge E_{\mathrm{break}},
\end{cases}
\label{eq:BPL}
\end{equation}
which captures changes in spectral slope due to radiative cooling, 
acceleration efficiency, or energy-dependent transport effects 
expected in evolved PWNe. For the \emph{Baseline}  and  \emph{CTAO-extended} analyses, electron
energies span the range $0.1$--$10^{4}$~TeV, while proton
energies extend from $1\,\mathrm{GeV}$ to $1\,\mathrm{PeV}$.
For the \emph{UHE-extended} analysis including LHAASO data, the
electron range is restricted to $0.1$--$300$~TeV, while the proton distribution extends from $1\,\mathrm{GeV}$ to $10\,\mathrm{PeV}$.
To investigate scenarios in which both leptonic and hadronic 
processes contribute, we construct a lepto-hadronic (LH) model in 
which the total $\gamma$-ray flux is
\begin{equation}
\Phi_{\mathrm{total}}(E_\gamma) = 
  \Phi_{\mathrm{IC}}(E_\gamma;\,N_e) + 
  \Phi_{\mathrm{PD}}(E_\gamma;\,N_p)\,.
\end{equation}
The underlying particle distributions are
\begin{equation}
N_e(E) = A_e\,\kappa_e(E), \qquad
N_p(E) = A_p\,\kappa_p(E),
\end{equation}
where $\kappa_e(E)$ and $\kappa_p(E)$ are BPL shape functions 
(Eq.~\ref{eq:BPL}) and $A_e$, $A_p$ are normalisation constants. 
The relative hadronic-to-leptonic contribution is controlled by the 
dimensionless ratio $\kappa_{pe} = A_p / A_e$. This formulation 
avoids arbitrary weighting at the photon level and enables a direct 
determination of the energy content of each particle population.

Model comparison is performed using the Bayesian Information Criterion (BIC; \citealt{Schwarz1978}),
\begin{equation}
\mathrm{BIC} = k\ln(n) - 2\ln(\mathcal{L}),
\end{equation}
where $k$ is the number of free parameters, $n$ the number of data 
points, and $\mathcal{L}$ the maximum likelihood. Models are ranked 
by the BIC difference relative to the purely leptonic (IC) scenario,
\begin{equation}
\Delta\mathrm{BIC}_{m} = 
  \mathrm{BIC}_{m} - \mathrm{BIC}_{\mathrm{IC}}\,.
\end{equation}
The statistically preferred model has the lowest BIC; accordingly, 
$\Delta\mathrm{BIC}_{m} < 0$ indicates that model~$m$ is favoured 
over the IC reference, while $\Delta\mathrm{BIC}_{m} > 0$ indicates 
it is disfavoured. Following the evidence scale of 
\citet{KassRaftery1995}, the strength of support is classified 
according to $|\Delta\mathrm{BIC}|$ as: not worth more than a bare 
mention ($<2$), positive ($2$--$6$), strong ($6$--$10$), or very 
strong ($\ge 10$).

\section{Results}
\label{sec:4}

We present the results of the broadband spectral modelling of 
HESS~J1825$-$137 under IC, PD, and LH emission scenarios. 
The analysis is organised in three stages: 
Section~\ref{sec:4.1} examines the \emph{Baseline} GeV--TeV dataset 
comprising \textit{Fermi}--LAT~\citep{Abdollahi2022,ballet2024}, H.E.S.S.~\citep{2019A&A...621A.116H}, HAWC~\citep{Albert2021}, and VERITAS~\citep{ABEYSEKARA2020102403}
observations; Section~\ref{sec:4.2} evaluates the 
discriminating power of future CTAO observations through 
prospective simulations; and Section~\ref{sec:4.3} 
incorporates LHAASO measurements~\citep{2024ApJS..271...25C} to assess the impact of UHE 
constraints on the inferred emission mechanisms. In each 
case, models are compared quantitatively using the BIC 
framework described in Section~\ref{sec:3}.

\subsection{Emission scenarios and statistical comparison using GeV--TeV data}
\label{sec:4.1}

The broadband $\gamma$-ray dataset considered in this section 
comprises observations from \textit{Fermi}--LAT~\citep{Abdollahi2022,ballet2024}, H.E.S.S.~\citep{2019A&A...621A.116H}, HAWC~\citep{Albert2021}, 
and VERITAS~\citep{ABEYSEKARA2020102403}, spanning the energy range from $\sim0.1$~GeV to 
$\sim160$~TeV. This \emph{Baseline} GeV--TeV dataset enables a direct 
comparison with previous studies and provides a reference framework 
for assessing the impact of additional UHE constraints 
(Sect.~\ref{sec:4.3}).

We fit the SED under three emission hypotheses: purely leptonic 
(IC), purely hadronic (PD), and combined lepto-hadronic (LH) 
scenarios. The statistical comparison, summarised in 
Table~\ref{tab:model-baseline}, shows that the IC model yields 
the lowest BIC and is therefore the statistically preferred 
description of the GeV--TeV data. The PD model is positively 
disfavoured ($\Delta\mathrm{BIC} = +3.07$), while all LH 
configurations incur a BIC penalty of $\Delta\mathrm{BIC} \geq 
+6.3$, reaching very strong evidence against for most values of 
$\kappa_{pe}$. The additional free parameter introduced by the LH 
model is not justified by the current data quality, indicating that 
the GeV--TeV observations alone do not require a hadronic 
contribution.

\begin{figure*}[h!]
    \centering
    \includegraphics[width=\textwidth]{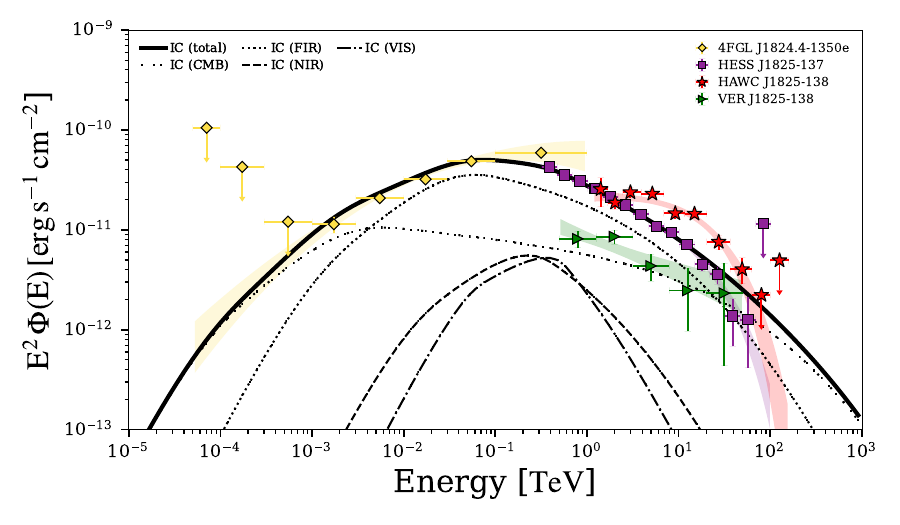}
    \caption{
    Spectral energy distribution of HESS~J1825--137 together with the best-fit inverse Compton (IC) model obtained from the baseline dataset (\textit{Fermi}-LAT,   H.E.S.S., HAWC, VERITAS).
    The solid curve shows the total IC emission, while the contributions from the cosmic 
microwave background (CMB), far-infrared (FIR), near-infrared (NIR), and optical/visible (VIS) photon fields are shown with different line styles.
    Gamma-ray data from \textit{Fermi}-LAT (diamond markers),   H.E.S.S. (square markers), HAWC (star markers), and VERITAS (triangle markers)  are overlaid. The shaded areas represent the statistical uncertainty of the spectral model fit. This IC model is used as the \emph{Baseline} scenario for comparison with
alternative emission models discussed in the text.
    }
    \label{fig:mode-baseline}
\end{figure*}

Figure~\ref{fig:mode-baseline} shows the best-fit IC 
model together with the multiwavelength data. The total IC emission 
is decomposed into contributions from four ambient photon fields. 
The FIR component dominates the $\gamma$-ray output across the 
$\sim0.1$--$10$~TeV range, consistent with the findings of ~\citet{2019A&A...621A.116H}, who identified the FIR field as the 
primary target for IC scattering in this source. The CMB 
contribution becomes comparable below $\sim10$~GeV and above 
$\sim50$~TeV, where Klein--Nishina suppression reduces the 
efficiency of scattering on higher-temperature fields. The NIR and 
VIS components provide subdominant but non-negligible contributions 
at intermediate energies.

The best-fit electron spectrum, described by a BPL 
(Eq.~\ref{eq:BPL}), has a low-energy index 
$\alpha_{1,e} = 1.35^{+0.49}_{-0.25}$, a high-energy index 
$\alpha_{2,e} = 3.18^{+0.02}_{-0.01}$, and a break energy 
$E_{\rm break,e} = 0.66^{+0.16}_{-0.06}$~TeV 
(Table~\ref{tab:model-baseline}). The steep post-break index is 
characteristic of strong synchrotron and IC cooling in an evolved 
PWN, and the spectral change $\Delta\alpha = \alpha_{2,e} - 
\alpha_{1,e} \approx 2.0$ exceeds the canonical cooling break of 
$\Delta\alpha = 1$ expected for continuous injection with purely 
radiative losses~\citep{1962SvA.....6..317K}, suggesting that 
energy-dependent escape or a time-varying injection history also 
shapes the electron distribution. The hard sub-break index 
$\alpha_{1,e} \approx 1.4$ is consistent with an uncooled 
population injected with a spectrum close to the theoretical 
prediction for relativistic shock acceleration in PWNe 
\citep{KirkSkjaeraasen2003, Sironi2011}.

The total energy in relativistic electrons, 
$W_e = 4.25 \times 10^{48}$~erg, is in good agreement with 
previous determinations for this source: 
\citet{2019A&A...621A.116H} obtained 
$W_e = (4.5 \pm 0.3) \times 10^{48}$~erg from H.E.S.S.\ data 
alone, \citet{2024MNRAS.528.2749C} found 
$W_e = (4.7 \pm 0.5) \times 10^{48}$~erg using a 3D transport 
model, and \citet{2024A&A...690A.116M} reported 
$W_e = (4.1 \pm 0.4) \times 10^{48}$~erg in a multi-zone IC+PD 
framework. The somewhat lower value of 
$W_e = (3.2 \pm 0.5) \times 10^{48}$~erg found by 
\citet{2020A&A...640A..76P} using \textit{Fermi}-LAT data alone 
reflects the narrower energy coverage of that analysis. The 
convergence of $W_e$ across independent studies employing different 
instruments, energy ranges, and modelling techniques confirms that the electron energy budget is robustly constrained at 
$W_e = 4.25 \times 10^{48}$~erg, representing 
$\sim 5\%$ of the total rotational energy released by 
PSR~J1826$-$1334 over its characteristic age.

In the purely hadronic scenario, broken power-law proton spectra 
can reproduce the TeV flux level with a total energy 
$W_p \sim \mathrm{few} \times 10^{48}$~erg, but the model 
systematically underpredicts the GeV emission measured by 
\textit{Fermi}-LAT, confirming that PD alone cannot account for the 
full broadband spectrum. This limitation reflects the kinematics of 
$\pi^0$ production, which imposes a low-energy threshold near 
$\sim100$~MeV in the $\gamma$-ray spectrum and cannot reproduce the 
hard GeV spectral index observed by \textit{Fermi}-LAT ~\citep{2020A&A...640A..76P}. LH models partially alleviate this 
tension by allowing the IC component to dominate at GeV energies, 
but the additional parameter $\kappa_{pe}$ is penalised by the BIC 
without a corresponding improvement in fit quality at the current 
sensitivity level. For clarity, only the best-fitting IC model is shown in Fig.~\ref{fig:mode-baseline}.

\begin{table}
\centering
\caption{Model comparison and best-fit parameters for 
HESS~J1825$-$137 using the GeV--TeV dataset. \textit{Upper panel:} 
BIC values for the IC, PD, and LH models. Boldface entries indicate the preferred model. \textit{Lower panel:} 
best-fit spectral parameters for the preferred IC model, with 
uncertainties corresponding to the 16th and 84th percentiles ($\sim\!1\sigma$) of the MCMC posterior distributions. The total  electron energy $W_e$ is a derived quantity integrated from the 
best-fit model. }
\begin{tabular}{lccccl}
\hline
\hline
Model & $\kappa_{pe}$ & $k$ & BIC & $\Delta\mathrm{BIC}$ 
  & Evidence$^a$ \\
\hline
\textbf{IC} & \textbf{--} & \textbf{3} & \textbf{454.47} 
  & \textbf{0.00} & \textbf{Reference (preferred)} \\
PD   & --      & 3 & 457.54 & $+3.07$  & Positive against \\
LH   & 0.001   & 4 & 467.21 & $+12.74$ & Very strong against \\
LH   & 0.01    & 4 & 465.20 & $+10.73$ & Very strong against \\
LH   & 0.1     & 4 & 465.54 & $+11.07$ & Very strong against \\
LH   & 1       & 4 & 462.48 & $+8.01$  & Strong against \\
LH   & 10      & 4 & 461.10 & $+6.63$  & Strong against \\
LH   & 100     & 4 & 460.78 & $+6.31$  & Strong against \\
LH   & 1000    & 4 & 465.70 & $+11.23$ & Very strong against \\
\hline
\hline
\multicolumn{2}{l}{Parameter} 
  & \multicolumn{4}{c}{IC (preferred)} \\
\hline
\multicolumn{2}{l}{$\log_{10}(N_{0,e}/\mathrm{eV}^{-1})$} 
  & \multicolumn{4}{c}{$36.25^{+0.11}_{-0.21}$} \\
\multicolumn{2}{l}{$\alpha_{1,e}$} 
  & \multicolumn{4}{c}{$1.35^{+0.49}_{-0.25}$} \\
\multicolumn{2}{l}{$\alpha_{2,e}$} 
  & \multicolumn{4}{c}{$3.18^{+0.02}_{-0.01}$} \\
\multicolumn{2}{l}{$E_{\mathrm{break},e}$ (TeV)} 
  & \multicolumn{4}{c}{$0.66^{+0.16}_{-0.06}$} \\
\multicolumn{2}{l}{$W_{e}$ ($10^{48}$~erg)} 
  & \multicolumn{4}{c}{$4.25$} \\
\hline
\end{tabular}
\vspace{1ex}
\raggedright\footnotesize{
$^a$ Evidence categories follow \citet{KassRaftery1995}. \\}
\label{tab:model-baseline}
\end{table}

\subsection{Impact of future CTAO observations}
\label{sec:4.2}

\begin{figure*}
    \centering
    \includegraphics[width=\textwidth]{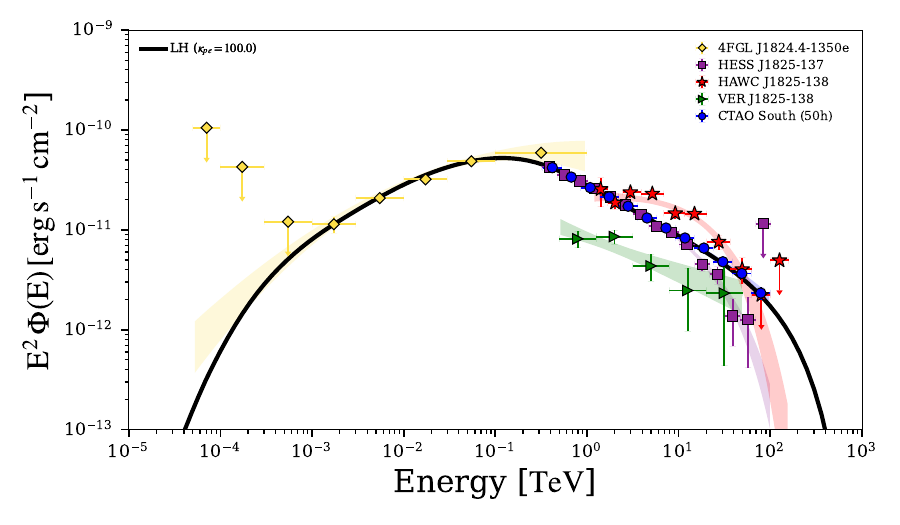}
    \caption{Spectral energy distribution of HESS\,J1825$-$137 together with the best-fit leptohadronic (LH) model with $\kappa_{pe}=100$ (solid curve), obtained from the joint fit of the \emph{Baseline} dataset (\textit{Fermi}-LAT,  H.E.S.S., HAWC,  VERITAS) augmented with the simulated 50\,h CTAO South observation. Gamma-ray data from \textit{Fermi}-LAT
(diamond markers),  H.E.S.S. (square markers),  HAWC (star markers),  VERITAS (triangle markers), and the simulated CTAO South flux points (circular markers)
are overlaid.  This LH configuration constitutes
the BIC-preferred scenario for the \emph{CTAO-extended} dataset (see
Table~\ref{tab:model-ctao-extended}).}
    \label{fig:model-ctao-extended}
\end{figure*}

To assess the diagnostic power of next-generation $\gamma$-ray 
observations, we augment the \emph{Baseline} GeV--TeV dataset with 
simulated CTAO observations produced using the Gammapy 
analysis toolkit \citep{Donath_2023} and the CTAO 
Southern Array IRFs (prod5 v0.1; 
\citealt{CTAIRFS2021}), assuming 50~h of on-source exposure. The 
CTAO spectral points, shown in 
Fig.~\ref{fig:model-ctao-extended}, densely sample the 
$\sim\!0.03$--$100$~TeV energy range with significantly 
reduced statistical uncertainties compared to the existing 
H.E.S.S.\ and VERITAS measurements, providing enhanced 
leverage in the multi-TeV regime where the predictions of 
leptonic and hadronic models diverge most strongly.

The inclusion of CTAO data qualitatively transforms the 
model comparison landscape relative to the \emph{Baseline} analysis. 
Whereas the GeV--TeV dataset alone yields a clear preference 
for the IC model, with all alternative scenarios disfavoured 
(Table~\ref{tab:model-baseline}), the \emph{CTAO-extended}  dataset 
reverses this conclusion 
(Table~\ref{tab:model-ctao-extended}). The purely leptonic 
IC model is no longer statistically preferred, the PD scenario 
achieves $\Delta\mathrm{BIC} = -27.97$, and the LH model with 
$\kappa_{pe} = 100$ yields the overall minimum BIC with 
$\Delta\mathrm{BIC} = -28.87$, both constituting very strong 
evidence in favour of a hadronic contribution. Several LH 
configurations spanning $\kappa_{pe} = 1$--$100$ are also 
very strongly favoured over IC, indicating that the preference 
for a hadronic component is robust against the specific choice 
of the proton-to-electron normalisation ratio within this range.

The best-fit spectral parameters of the preferred LH model 
are reported in the lower panel of 
Table~\ref{tab:model-ctao-extended}. 
The electron component retains spectral indices broadly consistent with
those obtained from the \emph{Baseline} IC-only fit (Table~\ref{tab:model-baseline}), although the preferred
leptohadronic solution favoured by the \emph{CTAO-extended} dataset
requires a substantially higher electron break energy.
This indicates that the low-energy gamma-ray emission remains
primarily constrained by the \textit{Fermi}-LAT and H.E.S.S.
data, while the CTAO measurements provide additional leverage
on the highest-energy portion of the particle distribution.
The corresponding electron energy budget is
$W_e = 0.02 \times 10^{48}$~erg, whereas the proton component
dominates the non-thermal energy content with
$W_p = 2.52 \times 10^{48}$~erg.

The physical origin of this shift lies in the precision with 
which CTAO constrains the spectral shape above 
$\sim\!10$~TeV. At these energies, the IC emission is 
increasingly suppressed by Klein--Nishina losses~\citep{1970RvMP...42..237B, 2005MNRAS.363..954M}, an effect 
that has been shown to be particularly relevant for PWNe 
emitting at UHE 
\citep{2022A&A...661A..72B}, and the 
predicted flux falls steeply with energy. The simulated CTAO 
data points in this regime, with their substantially reduced 
error bars, reveal a spectral hardening relative to the pure 
IC prediction that can be accommodated by a PD component 
arising from $pp$ interactions with the dense molecular 
material identified by \citet{2016MNRAS.458.2813V}. In the 
preferred LH scenario with $\kappa_{pe} = 100$, the IC 
component continues to dominate the emission below 
$\sim\!1$~TeV, while the hadronic contribution becomes 
progressively more important at higher energies and dominates 
above $\sim\!10$~TeV 
(Fig.~\ref{fig:model-ctao-extended}).

The BIC behaviour as a function of $\kappa_{pe}$ reveals a 
well-defined minimum in the range 
$\kappa_{pe} \sim 10$--$100$, with the statistical preference 
degrading sharply for both lower and higher values. LH models 
with $\kappa_{pe} \leq 0.1$, in which the hadronic 
component is subdominant, are very strongly disfavoured 
($\Delta\mathrm{BIC} \geq +26$), as the additional free 
parameter offers no improvement over the IC model in 
reproducing the high-energy data. Conversely, the 
$\kappa_{pe} = 1000$ configuration is also disfavoured 
($\Delta\mathrm{BIC} = +2.61$), indicating that an 
excessively dominant proton population overshoots the emission 
at intermediate energies where the IC component is well 
constrained by H.E.S.S.\ and CTAO measurements. This 
non-monotonic dependence of $\Delta\mathrm{BIC}$ on 
$\kappa_{pe}$ demonstrates that CTAO observations would not 
merely detect a hadronic contribution but could also 
constrain the relative energy content in relativistic 
electrons and protons.

It is worth comparing the \emph{CTAO-extended} results with those 
from previous analyses that considered only leptonic emission. 
The H.E.S.S.\ Collaboration analysis 
\citep{2019A&A...621A.116H}, based on $\sim\!400$~h of 
H.E.S.S.~I data, found a total electron energy of 
$W_e = (4.5 \pm 0.3) \times 10^{48}$~erg under a purely IC 
framework. The preferred CTAO leptohadronic model requires a proton
energy budget of $W_p = 2.52 \times 10^{48}$~erg. The
improvement in statistical preference relative to the purely IC
scenario indicates that the precise CTAO measurements in the
multi-TeV regime are sensitive to an additional hadronic
component, which becomes increasingly important above
$\sim10$~TeV and dominates the emission at the highest
energies.
 The proton energy budget 
is consistent with the value 
$W_p = (1.8 \pm 0.7) \times 10^{48}$~erg reported by 
\citet{2024A&A...690A.116M} in their multi-zone IC+PD model, 
although our one-zone analysis yields a somewhat higher 
central value, as expected when spatial segregation between 
the electron and proton populations is not modelled explicitly.

These results demonstrate that a 50~h CTAO observation of 
HESS~J1825$-$137 would provide sufficient sensitivity to 
unambiguously break the degeneracy among emission scenarios 
that persists with current instruments. The transition from a 
\emph{Baseline} IC preference to a very strong LH preference 
underscores the critical importance of precise spectral 
measurements in the $\sim\!10$--$100$~TeV range, where 
Klein--Nishina suppression of IC emission creates a spectral 
window uniquely sensitive to hadronic processes. Such 
observations would establish whether the molecular material 
adjacent to the nebula \citep{2016MNRAS.458.2813V} serves as 
an effective target for hadronic $\gamma$-ray production and 
would place direct constraints on the proton energy budget, 
with implications for the role of evolved PWNe as sources of 
hadronic cosmic rays.

\begin{table}
\centering
\caption{Same as Table~\ref{tab:model-baseline} but for the 
\emph{Baseline} dataset including simulated CTAO observations. Boldface entries indicate the preferred model, corresponding to the leptohadronic (LH) scenario with
$\kappa_{pe} = 100$.}
\label{tab:model-ctao-extended}
\begin{tabular}{lccccl}
\hline
\hline
Model & $\kappa_{pe}$ & $k$ & BIC & $\Delta\mathrm{BIC}$ 
  & Evidence$^a$ \\
\hline
IC    & --      & 3 & 527.34 & $0.00$    & Ref. \\
PD    & --      & 3 & 499.37 & $-27.97$  & VS for \\
LH    & 0.001   & 4 & 622.59 & $+95.25$  & VS against \\
LH    & 0.01    & 4 & 610.94 & $+83.60$  & VS against \\
LH    & 0.1     & 4 & 553.68 & $+26.34$  & VS against \\
LH    & 1       & 4 & 517.15 & $-10.19$  & VS for \\
LH    & 10      & 4 & 502.02 & $-25.32$  & VS for \\
\textbf{LH} & \textbf{100} & \textbf{4} 
  & \textbf{498.47} & $\boldsymbol{-28.87}$ 
  & \textbf{VS for} \\
LH    & 1000    & 4 & 529.95 & $+2.61$   
  & P against \\
\hline
\hline
\multicolumn{2}{l}{Parameter} 
  & \multicolumn{4}{c}{LH, $\kappa_{pe} = 100$ (preferred)} \\
\hline
\multicolumn{6}{c}{\textit{Electron component}} \\
\multicolumn{2}{l}{$\log_{10}(N_{0,e}/\mathrm{eV}^{-1})$} 
  & \multicolumn{4}{c}{$33.57^{+0.05}_{-0.06}$} \\
\multicolumn{2}{l}{$\alpha_{1,e}$} 
  & \multicolumn{4}{c}{$1.83^{+0.42}_{-0.49}$} \\
\multicolumn{2}{l}{$\alpha_{2,e}$} 
  & \multicolumn{4}{c}{$3.17^{+0.53}_{-0.36}$} \\
\multicolumn{2}{l}{$E_{\mathrm{break},e}$ (TeV)} 
  & \multicolumn{4}{c}{$1.77^{+0.03}_{-0.54}$} \\
\multicolumn{2}{l}{$W_{e}$ ($10^{48}$~erg)} 
  & \multicolumn{4}{c}{$0.02$} \\
\hline
\multicolumn{6}{c}{\textit{Proton component}} \\
\multicolumn{2}{l}{$\log_{10}(N_{0,p}/\mathrm{eV}^{-1})$} 
  & \multicolumn{4}{c}{$35.57^{+0.05}_{-0.06}$} \\
\multicolumn{2}{l}{$\alpha_{1,p}$} 
  & \multicolumn{4}{c}{$1.50^{+0.10}_{-0.08}$} \\
\multicolumn{2}{l}{$\alpha_{2,p}$} 
  & \multicolumn{4}{c}{$2.56^{+0.02}_{-0.01}$} \\
\multicolumn{2}{l}{$E_{\mathrm{break},p}$ (TeV)} 
  & \multicolumn{4}{c}{$1.34^{+0.26}_{-0.15}$} \\
\multicolumn{2}{l}{$W_{p}$ ($10^{48}$~erg)} 
  & \multicolumn{4}{c}{$2.52$} \\
\hline
\end{tabular}

\vspace{1ex}
\raggedright\footnotesize{
$^a$P\,=\,positive,  
VS\,=\,very strong; see 
\citet{KassRaftery1995}. \\}
\end{table}

\subsection{Impact of UHE constraints from LHAASO}
\label{sec:4.3}

We now extend the \emph{Baseline} GeV--TeV analysis by incorporating 
the UHE measurements provided by LHAASO 
\citep{2024ApJS..271...25C}, resulting in a combined GeV--PeV 
dataset spanning from $\sim\!0.1$~GeV to $\sim\!630$~TeV. 
This extended energy coverage allows us to probe the particle 
populations responsible for the $\gamma$-ray emission of 
HESS~J1825$-$137 up to energies approaching the PeV scale, a 
regime where the predictions of leptonic and hadronic models 
diverge most dramatically.

Before presenting the spectral modelling results, it is 
important to discuss the spatial context of the LHAASO 
measurements. As illustrated in Fig.~\ref{fig:skymap}, the 
LHAASO source J1825$-$1326~\citep{2021Natur.594...33C, 
2024ApJS..271...25C} has a considerably larger angular 
extent than the individual TeV sources in the region and 
overlaps spatially with multiple $\gamma$-ray emitters, 
including HESS~J1825$-$137, the hard-spectrum source 
HESS~J1826$-$130, and the HAWC component 
eHWC~J1825$-$134 
\citep[see also][]{Costa2024}. Owing to the limited angular resolution of 
air-shower arrays at these energies, the LHAASO flux cannot 
be unambiguously attributed to a single astrophysical 
counterpart. In particular,~\citet{2024ApJS..271...25C} 
report HESS~J1825$-$137 as the closest known counterpart to 
1LHAASO~J1825$-$1418 (separation $0.56^{\circ}$) and 
1LHAASO~J1825$-$1337u ($0.15^{\circ}$), while 
HESS~J1826$-$130, associated with the Eel PWN 
(G18.5$-$0.4) and PSR~J1826$-$1256~\citep{Duvidovich2019, Anguner2017}, lies 
within the same extended emission region. Similarly, the HAWC 
decomposition by \citet{2021ApJ...907L..30A} identified a 
distinct hard source, HAWC~J1825$-$134, whose emission 
extends beyond 200~TeV without a cutoff and is spatially 
coincident with dense molecular material rather than the PWN 
itself. This spatial complexity implies that the LHAASO flux 
points used in our analysis should be regarded as upper limits 
on the contribution of HESS~J1825$-$137 alone, and the model 
comparison results interpreted accordingly.

With this caveat in mind, we repeat the spectral fitting for 
IC, PD, and LH scenarios using the GeV--PeV dataset. The 
statistical comparison based on the BIC is summarised in 
Table~\ref{tab:model-uhe-extended}. The inclusion of 
LHAASO data substantially alters the model preference 
relative to the \emph{Baseline} analysis. The PD scenario now 
achieves the lowest BIC, with 
$\Delta\mathrm{BIC} = -7.89$ relative to the IC reference, 
constituting strong evidence in favour of a hadronic 
interpretation. Among the LH configurations, the model with 
$\kappa_{pe} = 1$ is also strongly favoured 
($\Delta\mathrm{BIC} = -6.90$), while $\kappa_{pe} = 0.1$ 
and $\kappa_{pe} = 10$ show positive evidence for a hadronic 
contribution. Higher values of $\kappa_{pe}$ yield 
inconclusive results ($|\Delta\mathrm{BIC}| < 2$ for 
$\kappa_{pe} = 100$ and $1000$), and LH models with very low 
hadronic fractions ($\kappa_{pe} \leq 0.01$) are strongly to 
very strongly disfavoured, as the additional parameter 
provides no benefit when the proton component is negligible.

The best-fit PD proton spectrum, reported in the lower panel 
of Table~\ref{tab:model-uhe-extended}, is described by a 
broken power law with a hard low-energy index 
$\alpha_{1,p} = 1.43^{+0.12}_{-0.13}$, a softer 
high-energy index 
$\alpha_{2,p} = 2.55^{+0.01}_{-0.01}$, and a break energy 
$E_{\mathrm{break},p} = 1.21^{+0.27}_{-0.16}$~TeV. The 
total proton energy is 
$W_p = 2.48 \times 10^{48}$~erg. This energy budget is 
energetically viable: adopting a spherical molecular cloud of 
radius $R_{\mathrm{MC}} \sim 18$~pc with an average hydrogen 
density $n_{\mathrm{H}} \simeq 6 \times 10^{2}$~cm$^{-3}$ 
\citep{2016MNRAS.458.2813V}, the $pp$ interaction timescale 
is $t_{pp} \simeq (n_{\mathrm{H}} \, \sigma_{pp} \, c)^{-1} 
\sim 10^{5}$~yr, implying that only a fraction of the proton 
energy has been radiated over the $\sim\!20$--$40$~kyr age of 
the system. The inferred $W_p$ is also consistent with the 
value $W_p = (1.8 \pm 0.7) \times 10^{48}$~erg obtained by 
\citet{2024A&A...690A.116M} in their multi-zone IC+PD 
framework, with the somewhat higher central value in our 
analysis reflecting the broader energy range probed by the 
LHAASO data.

\begin{figure*}
    \centering
    \includegraphics[width=\textwidth]{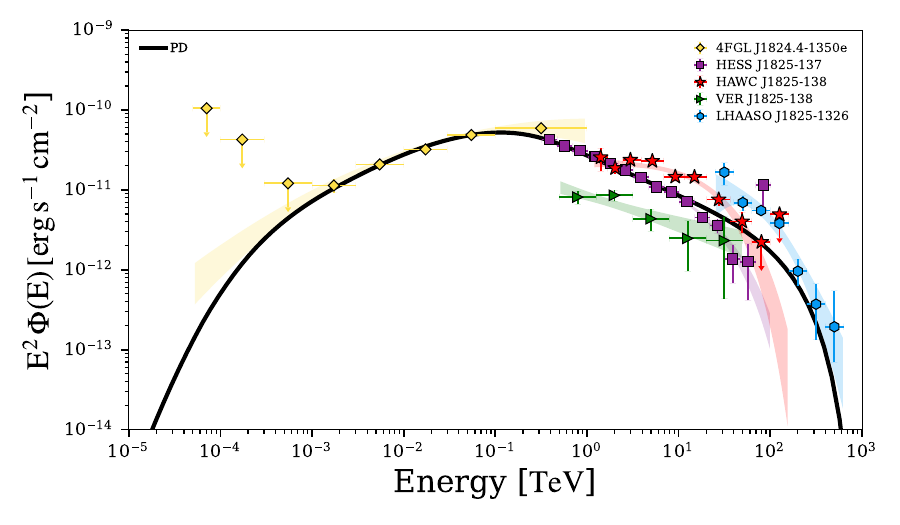}
    \caption{Spectral energy distribution of HESS\,J1825$-$137 together with the best-fit purely hadronic pion-decay (PD) model obtained from the \emph{UHE-extended} dataset (\emph{Baseline} + LHAASO). The solid curve shows the
$\pi^{0}$-decay gamma-ray emission produced by $pp$ interactions of the
relativistic proton population with the dense molecular material of
mean density $n_{\rm H}\simeq 6\times 10^{2}\,{\rm cm^{-3}}$ adopted
from \citet{Voisin2016}. Gamma-ray data from \textit{Fermi}-LAT (yellow
diamonds),   H.E.S.S. (purple markers), HAWC (orange markers), VERITAS (green markers) and LHAASO\,J1825$-$1326 (hexagon markers,
\citealt{Cao2021b}) are overlaid.  This PD
configuration is the BIC-preferred scenario for the \emph{UHE-extended} dataset
(Table~\ref{tab:model-uhe-extended}).
    }
    \label{fig:mode-uhe-extended}
\end{figure*}

Figure~\ref{fig:mode-uhe-extended} shows the best-fit 
PD model together with the GeV--PeV dataset. The PD
spectrum reproduces the TeV emission measured by H.E.S.S.\ 
and HAWC and provides a reasonable description of the LHAASO 
flux points up to $\sim\!200$~TeV. At the highest LHAASO 
energies ($E_{\gamma} \gtrsim 300$~TeV), the model curve 
begins to fall below the data, which may reflect 
contributions from spatially coincident sources not accounted 
for in the one-zone framework, most notably 
HAWC~J1825$-$134 and HESS~J1826$-$130, both of which are 
hard-spectrum emitters in this energy range 
\citep{2021ApJ...907L..30A, 2018A&A...612A...2H}. At GeV 
energies, the PD model passes through the 
\textit{Fermi}-LAT data points, unlike in the \emph{Baseline} 
analysis where the PD scenario systematically underpredicted 
the GeV flux. This improvement arises because the LHAASO 
constraints force a harder sub-break proton index 
($\alpha_{1,p} = 1.43^{+0.12}_{-0.13}$), which produces a harder $\pi^{0}$ 
spectrum that extends to lower $\gamma$-ray energies.

\begin{figure}
    \centering
    \includegraphics[width=0.5\textwidth]{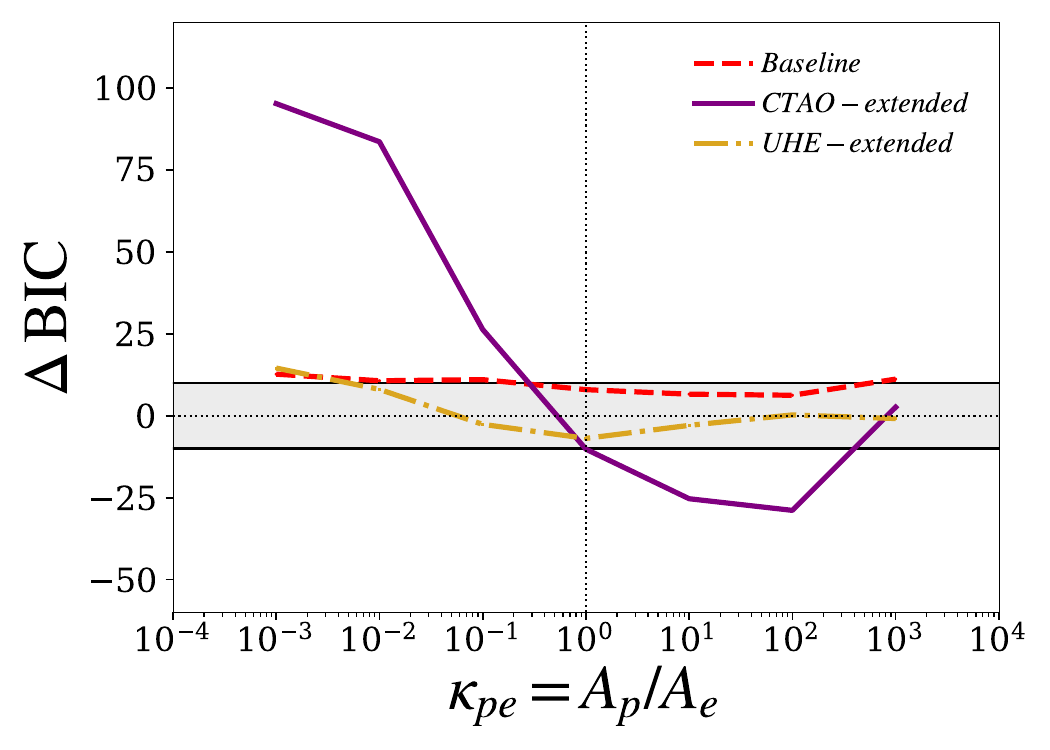}
    \caption{Difference in Bayesian Information Criterion,
$\Delta \mathrm{BIC}_m = \mathrm{BIC}_m - \mathrm{BIC}_{\mathrm{IC}}$, where $m$ denotes a given leptohadronic model,
as a function of the proton-to-electron normalization ratio
$\kappa_{pe} = A_p/A_e$ for leptohadronic models.
The dashed, solid, and dash-dotted curves correspond to the \emph{Baseline}, \emph{CTAO-extended}, and \emph{UHE-extended} datasets, respectively.
Negative values of $\Delta \mathrm{BIC}$ indicate a statistical preference for the tested model
with respect to the IC-only scenario.
The vertical dotted line marks $\kappa_{pe}=1$, corresponding to equal proton and electron normalization. The shaded region corresponds to the interval
$-10 < \Delta \mathrm{BIC} < 10$. Values of $|\Delta \mathrm{BIC}| \gtrsim 10$ correspond to strong evidence.
}\label{fig:delta_bic_kappa}
\end{figure}

The contrasting model preferences across the three dataset 
configurations are illustrated in 
Fig.~\ref{fig:delta_bic_kappa}, which shows 
$\Delta\mathrm{BIC}_{m}$ as a function of $\kappa_{pe}$. The 
\emph{Baseline} data (red dashed) yield positive 
$\Delta\mathrm{BIC}$ across all $\kappa_{pe}$, confirming 
that no hadronic contribution is required. The \emph{UHE-extended}
dataset (orange dot-dashed) crosses zero near 
$\kappa_{pe} \sim 0.1$ and reaches a broad minimum around 
$\kappa_{pe} \sim 1$, but returns to 
$\Delta\mathrm{BIC} \approx 0$ for $\kappa_{pe} \geq 100$, 
indicating that the hadronic preference is confined to a 
narrow range of mixing ratios. The \emph{CTAO-extended} dataset 
(purple solid) exhibits the most dramatic behaviour, dropping 
monotonically from $\Delta\mathrm{BIC} \approx +95\%$ at 
$\kappa_{pe} = 10^{-3}$ to $\approx -30\%$ at 
$\kappa_{pe} \sim 100$, demonstrating that future CTAO 
observations would provide unambiguous discrimination among 
emission scenarios across the full range of $\kappa_{pe}$.

It is nevertheless essential to interpret the PD preference 
with appropriate caution. The shift from IC to PD as the 
statistically preferred model depends critically on the 
assumption that the LHAASO flux originates predominantly from 
HESS~J1825$-$137. As discussed above and shown in 
Fig.~\ref{fig:skymap}, the LHAASO emission 
region encompasses multiple potential contributors. If a 
significant fraction of the UHE flux arises from 
HESS~J1826$-$130 or from the distinct HAWC component 
J1825$-$134, which~\citet{2021ApJ...907L..30A} associates 
with dense molecular material rather than the PWN, then the 
effective UHE flux attributable to HESS~J1825$-$137 would be 
lower, potentially weakening or eliminating the statistical 
preference for the PD model. Future observations with 
improved angular resolution, particularly by CTAO~\citep{2019scta.book.....C} and next-generation wide-field arrays, 
will be essential to disentangle the UHE contributions from 
individual sources in this crowded region and to establish 
whether the hadronic preference identified here reflects a 
genuine property of the PWN or a superposition of unresolved 
emission components.

\begin{table}
\centering
\caption{Same as Table~\ref{tab:model-baseline} but for the 
\emph{UHE-extended} dataset including LHAASO observations. Boldface entries indicate the preferred model, corresponding to the PD scenario.}
\label{tab:model-uhe-extended}
\begin{tabular}{lccccl}
\hline
\hline
Model & $\kappa_{pe}$ & $k$ & BIC & $\Delta\mathrm{BIC}$ 
  & Evidence$^a$ \\
\hline
IC    & --      & 3 & 528.59 & $0.00$    & Ref. \\
\textbf{PD} & \textbf{--} & \textbf{3} & \textbf{520.70} 
  & $\boldsymbol{-7.89}$ & \textbf{S for} \\
LH    & 0.001   & 4 & 543.14 & $+14.55$  & VS against \\
LH    & 0.01    & 4 & 536.63 & $+8.04$   & S against \\
LH    & 0.1     & 4 & 525.97 & $-2.62$   & P for \\
LH    & 1       & 4 & 521.69 & $-6.90$   & S for \\
LH    & 10      & 4 & 525.66 & $-2.93$   & P for \\
LH    & 100     & 4 & 528.85 & $+0.26$   & I \\
LH    & 1000    & 4 & 527.74 & $-0.85$   & I \\
\hline
\hline
\multicolumn{2}{l}{Parameter} 
  & \multicolumn{4}{c}{PD (preferred)} \\
\hline
\multicolumn{2}{l}{$\log_{10}(N_{0,p}/\mathrm{eV}^{-1})$} 
  & \multicolumn{4}{c}{$35.59^{+0.07}_{-0.07}$} \\
\multicolumn{2}{l}{$\alpha_{1,p}$} 
  & \multicolumn{4}{c}{$1.43^{+0.12}_{-0.13}$} \\
\multicolumn{2}{l}{$\alpha_{2,p}$} 
  & \multicolumn{4}{c}{$2.55^{+0.01}_{-0.01}$} \\
\multicolumn{2}{l}{$E_{\mathrm{break},p}$ (TeV)} 
  & \multicolumn{4}{c}{$1.21^{+0.27}_{-0.16}$} \\
\multicolumn{2}{l}{$W_{p}$ ($10^{48}$~erg)} 
  & \multicolumn{4}{c}{$2.48$} \\
\hline
\end{tabular}

\vspace{1ex}
\raggedright\footnotesize{
$^a$I\,=\,inconclusive, P\,=\,positive, S\,=\,strong, 
VS\,=\,very strong;  see 
\citet{KassRaftery1995}. \\
}
\end{table}

\section{Discussion}
\label{sec:5}

The broadband spectral modelling presented in 
Sections~\ref{sec:3} and~\ref{sec:4} provides a coherent 
picture of the particle populations responsible for the 
$\gamma$-ray emission of HESS~J1825$-$137. For the \emph{Baseline} GeV--TeV dataset,  the electron energy
distribution is well described by a broken power law with
$\alpha_{1,e}=1.35^{+0.49}_{-0.25}$,
$\alpha_{2,e}=3.18^{+0.02}_{-0.01}$, and
$E_{\rm break,e}=0.66^{+0.16}_{-0.06}$~TeV.
The corresponding electron energy budget,
$W_e = 4.25 \times 10^{48}$~erg, is in good agreement with
previous determinations from H.E.S.S.
\citep{2019A&A...621A.116H}, 3D transport
modelling~\citep{2024MNRAS.528.2749C},
and multi-zone radiative models~\citep{2024A&A...690A.116M}, confirming that 
$W_e$ is robustly constrained at $\sim 6\%$ of the 
total rotational energy of PSR~J1826$-$1334 and compatible 
with values derived for other extended PWNe such as 
G21.5$-$0.9 and 
Vela~X~\citep{2010ApJ...724..572M, 2011A&A...525A.154S, 
2019JCAP...01..046B}. The super-canonical spectral 
steepening $\Delta\alpha \approx 2.0$, exceeding the 
$\Delta\alpha = 1$ predicted by 
\citet{1962SvA.....6..317K}, reflects the spatially-averaged 
nature of our one-zone model, which superposes recently injected and radiatively cooled electron populations distributed
across the $\sim 100$~pc nebula, compounded by 
energy-dependent transport ($R \propto E^{-0.29}$; 
\citealt{2019A&A...621A.116H}).

The most notable result of this work is the 
dataset-dependent shift in model preference. The \emph{Baseline} 
GeV--TeV data favour pure IC emission, with all alternatives 
disfavoured (Table~\ref{tab:model-baseline}). 
Simulated CTAO data reverse this conclusion: the LH model 
with $\kappa_{pe} = 100$ becomes very strongly preferred 
($\Delta\mathrm{BIC} = -28.87$), driven by the precision 
above $\sim 10$~TeV where Klein--Nishina 
suppression~\citep{1970RvMP...42..237B, 
2005MNRAS.363..954M, 2022A&A...661A..72B} creates a spectral 
window sensitive to hadronic processes. When LHAASO data are 
included, the PD model achieves the lowest BIC 
($\Delta\mathrm{BIC} = -7.89$), though the preference is 
confined to $\kappa_{pe} \sim 0.1$--$10$ 
(Fig.~\ref{fig:delta_bic_kappa}). The inferred proton energy 
$W_p \approx 2.5 \times 10^{48}$~erg is consistent 
with the multi-zone estimate 
of~\citet{2024A&A...690A.116M} and energetically viable 
given the dense molecular material identified 
by~\citet{2016MNRAS.458.2813V} ($n_{\rm H} \simeq 
6 \times 10^{2}$~cm$^{-3}$, 
$t_{pp} \sim 10^{5}$~yr).

Several caveats should be noted. The one-zone approximation 
does not capture the spatially resolved spectral evolution 
revealed by deep H.E.S.S. observations and multi-zone 
models~\citep{2019A&A...621A.116H, 2024MNRAS.528.2749C, 
2024A&A...690A.116M}; the effective parameters represent 
population-averaged quantities. The PD preference in the 
\emph{UHE-extended} analysis depends on the assumption that the 
LHAASO flux originates predominantly from 
HESS~J1825$-$137, whereas the emission region encompasses 
multiple contributors including HESS~J1826$-$130 and 
HAWC~J1825$-$134~\citep{2021ApJ...907L..30A}. If a 
significant fraction of the UHE flux arises from these 
neighbours, the hadronic preference could weaken or 
disappear. The CTAO projections rely on current IRFs 
(prod5 v0.1) and assume a single spectral component; the 
actual discriminating power will depend on the realised 
sensitivity and ability to spatially resolve the region.

PWNe are conventionally regarded as paradigmatic leptonic
$\gamma$-ray emitters. Their bolometric output is
overwhelmingly attributed to synchrotron and IC
radiation from the relativistic pair plasma launched at the
pulsar wind termination shock, and the leptonic energy
budget recovered here for HESS~J1825$-$137
($W_e = 4.25 \times 10^{48}$~erg, i.e.\ $\sim 6\%$ of
the rotational energy released by PSR~J1826$-$1334 over its
characteristic age) is fully consistent with that picture.
The \emph{CTAO-extended} and \emph{UHE-extended} analyses of
Sections~\ref{sec:4.2} and~\ref{sec:4.3} nevertheless favour
the inclusion of a hadronic component carrying
$W_p \approx 2.5 \times 10^{48}$~erg, comparable in
order of magnitude to $W_e$. The broad-band spectrum of an
evolved PWN cannot therefore be reproduced under a purely
leptonic assumption, once the multi-TeV and UHE regimes are
precisely measured.

The physical origin of the inferred hadrons can be ascribed
to two complementary scenarios that are not mutually exclusive. In the first, hadrons are accelerated within the
PWN itself, for instance, at the termination shock by an
ion-loaded pulsar wind, or via secondary processes triggered
by the relativistic electrons interacting with the
surrounding magnetized plasma. Evolved PWNe would then not
be purely leptonic systems and would contribute, at some
level, to the Galactic cosmic-ray proton census. Even
moderate per-source energetics, multiplied by the Galactic
population of evolved PWNe with characteristic ages of tens
of kyr, could amount to a non-negligible contribution to the
cosmic-ray budget below the knee. In the second scenario,
suggested by the CO(1--0) and CS(1--0) observations
of~\citet{2016MNRAS.458.2813V} that motivate our hadronic
target, the inferred protons are not accelerated by the PWN
itself, but correspond to relativistic particles escaping the
parent SNR (and possibly other surrounding accelerators)
which subsequently illuminate the dense molecular material
adjacent to the nebula. The BIC-preferred hadronic component
is then a tracer of the integrated cosmic-ray production
history of this complex region rather than a property of the
PWN in isolation. At the level of the spatially-averaged
one-zone analysis performed here, the two pictures are
degenerate. Disentangling them requires the improved angular
resolution that CTAO will provide, which would allow the
$\gamma$-ray emission spatially correlated with the
molecular clumps to be separated from the diffuse PWN
emission centered on PSR~J1826$-$1334, as already noted in
the discussion of Fig.~\ref{fig:skymap} and the LHAASO
source-confusion issue.

The inferred hadronic energetics also carry a temporal
implication. With $W_p \approx 2.5 \times
10^{48}$~erg, a target density
$n_{\rm H} \simeq 6 \times 10^{2}$~cm$^{-3}$ and a $pp$
interaction timescale $t_{pp} \sim 10^{5}$~yr, only a
fraction of the available proton energy has been radiated
over the $\sim 20$-$40$~kyr age of the system. The region
therefore remains an active site of hadronic $\gamma$-ray
production, and is a prime target for the next generation of
observatories sensitive above $\sim 100$~TeV (CTAO South,
SWGO, and continued LHAASO operations), which will be in a
position to test the scenarios outlined above and to
quantify the contribution of evolved PWNe to the Galactic
cosmic-ray proton population.

Table~\ref{tab:literature_comparison} places our results in 
the context of previous analyses. The present work extends 
these studies by providing the broadest multi-instrument 
energy coverage to date and the first application of 
BIC-based model selection to this source. The consistency of 
$W_e$ for the
current GeV--TeV dataset underscores the robustness of the 
electron energy budget as a fundamental property of this PWN, 
while the hadronic budget 
$W_p \approx 2.5 \times 10^{48}$~erg, constrained here 
for the first time with formal model selection, confirms that 
the dense environment of HESS~J1825$-$137 can serve as an 
effective target for hadronic $\gamma$-ray production.

\begin{table*}
\centering
\caption{Comparison of broadband spectral analyses of 
HESS~J1825$-$137. Energy ranges reflect the effective 
combined coverage. Dashes indicate unconstrained parameters.}
\label{tab:literature_comparison}
\begin{tabular}{lccccc}
\hline\hline
Study & Instruments & $E$ range & Model 
& $W_e$ & $W_p$ \\
& & (TeV) & 
& ($10^{48}$\,erg) & ($10^{48}$\,erg) \\
\hline
H.E.S.S.\,(2019)$^a$         
& H.E.S.S.              
& 0.3--30         
& IC      
& $4.5 \pm 0.3$ 
& --- \\

Principe\,(2020)$^b$         
& \textit{Fermi}-LAT    
& 0.001--0.1      
& IC      
& $3.2 \pm 0.5$ 
& --- \\

Martin\,(2024)$^c$           
& H.E.S.S.+\textit{Fermi}  
& 0.1--30         
& IC+PD   
& $4.1 \pm 0.4$ 
& $1.8 \pm 0.7$ \\

Collins\,(2024)$^d$          
& H.E.S.S.              
& 0.3--30         
& IC      
& $4.7 \pm 0.5$ 
& --- \\
\hline

\textbf{Baseline}$^{\dagger}$ 
& Multi-instrument$^e$  
& 0.1--160        
& IC 
& 4.25 
& --- \\

\textbf{CTAO-ext.}$^{\dagger}$ 
& Baseline+CTAO$^f$     
& 0.1--160        
& LH 
& 0.02
& 2.52 \\

\textbf{UHE-ext.}$^{\dagger}$ 
& Baseline+LHAASO       
& 0.1--630        
& PD 
& ---
& 2.48 \\
\hline
\end{tabular}

\vspace{1ex}
\raggedright\footnotesize{
$^{\dagger}$Boldface entries indicate the dataset configurations analysed in this work; all configurations use BIC/MCMC model selection. \\
$^a$\citet{2019A&A...621A.116H}. \;
$^b$\citet{2020A&A...640A..76P}. \;
$^c$\citet{2024A&A...690A.116M}. \;
$^d$\citet{2024MNRAS.528.2749C}. \\
$^e$\textit{Fermi}-LAT + H.E.S.S. + HAWC + VERITAS. \;
$^f$Simulated 50\,h CTAO South (prod5 v0.1). \\
}
\end{table*}

\section{Conclusion}
\label{sec:6}

We have presented a comprehensive spectral analysis of the 
$\gamma$-ray emission from HESS~J1825$-$137, combining data 
from \textit{Fermi}--LAT~\citep{Abdollahi2022,ballet2024}, H.E.S.S.~\citep{2019A&A...621A.116H}, HAWC~\citep{Albert2021}, and VERITAS~\citep{ABEYSEKARA2020102403} with simulated 
CTAO observations and LHAASO measurements~\citep{2024ApJS..271...25C}. Using \textsc{Naima}~\citep{naima} with MCMC parameter estimation and BIC 
model selection, we have quantitatively compared purely 
leptonic, purely hadronic, and lepto-hadronic
emission scenarios across three dataset configurations.

The \emph{Baseline} GeV--TeV dataset 
($\sim 0.1$~GeV--$160$~TeV) favours the IC model, with all 
hadronic alternatives statistically disfavoured 
($\Delta\mathrm{BIC} \geq +3$). The best-fit electron 
spectrum is described by a broken power law with 
$\alpha_{1,e} = 1.35^{+0.49}_{-0.25}$, $\alpha_{2,e} = 3.18^{+0.02}_{-0.01}$, 
$E_{\rm break,e} = 0.66^{+0.16}_{-0.06}$~TeV, and a total energy 
$W_e =4.25 \times 10^{48}$~erg, in excellent agreement 
with previous studies. Simulated 50\,h CTAO South 
observations dramatically reverse this preference: the LH 
scenario with $\kappa_{pe} = 100$ becomes very strongly 
preferred ($\Delta\mathrm{BIC} = -28.87$), demonstrating 
that CTAO would provide definitive discrimination between 
emission scenarios by precisely constraining the spectral 
shape above $\sim 10$~TeV, where Klein--Nishina suppression 
of IC emission opens a window to hadronic processes. The 
inclusion of LHAASO data ($\sim 0.1$~GeV--$630$~TeV) shifts 
the preference toward the PD model 
($\Delta\mathrm{BIC} = -7.89$), with LH models at 
$\kappa_{pe} \sim 0.1$--$10$ also favoured, although this 
result must be interpreted with caution given the spatial 
confusion in the LHAASO source region, which encompasses 
contributions from HESS~J1826$-$130 and 
HAWC~J1825$-$134. The inferred proton energy budget, 
$W_p \approx 2.5 \times 10^{48}$~erg, is energetically consistent with 
$pp$ interactions in the dense molecular material 
($n_{\rm H} \simeq 6 \times 10^{2}$~cm$^{-3}$) adjacent 
to the nebula.

Future CTAO observations with improved angular resolution 
will be essential to spatially disentangle the emission from 
HESS~J1825$-$137 and its neighbours, and to establish 
whether the hadronic component identified in this work 
reflects a genuine property of the PWN environment~\citep{2019scta.book.....C}. Combined 
with next-generation wide-field arrays, such observations 
will place definitive constraints on the role of evolved 
PWNe as sources of hadronic cosmic rays in the Galaxy.

\section*{Acknowledgements}

We sincerely thank the referee for their thoughtful feedback and valuable suggestions, which have enhanced the clarity and scientific rigor of this work. The authors acknowledge the financial support of the NAPI “Fenômenos Extremos do Universo” of Fundação de Apoio à Ciência, Tecnologia e Inovação do Paraná. Araucária Foundation (grant Nos. 698/2022 and 721/2022) and FAPESP (grant No. 2021/01089-1). The authors acknowledge the National Laboratory for Scientific Computing (LNCC/MCTI, Brazil) for providing HPC resources of the SDumont supercomputer, which have contributed to the research results reported in this paper. URL:  \href{https://sdumont.lncc.br}{https://sdumont.lncc.br}.  The research also employed \textsc{Gammapy}, a Python package developed by the community for TeV gamma-ray astronomy \citep{Donath_2023}, accessible at \href{https://www.gammapy.org}{https://www.gammapy.org}. In addition, we used the instrument response functions for the Cherenkov Telescope Array (CTA) provided by the CTA Consortium and Observatory. For detailed information about these instrument response functions, please refer to \href{https://www.ctao.org/for-scientists/performance/}{https://www.ctao.org/for-scientists/performance/} (version prod5 v0.1; \citealt{CTAIRFS2021}).

\section*{Data Availability}

The data underlying this article are publicly available.
\textit{Fermi}-LAT data can be accessed via the Fermi Science
Support Center (\url{https://fermi.gsfc.nasa.gov/ssc/}),
while H.E.S.S., HAWC, and VERITAS data are available through
the respective collaborations' publications. LHAASO data are
available from published results. The CTAO dataset used in
this work is simulated and was generated using the publicly
available Gammapy software package and CTAO instrument
response functions. The datasets generated
during this study are available from the corresponding author
upon reasonable request.



\bibliographystyle{mnras}
\bibliography{bibliograph} 


\bsp	
\label{lastpage}
\end{document}